%% file: main.tex
\documentclass[lettersize,journal]{IEEEtran}
\usepackage{amsmath,amsfonts}
\usepackage{algorithmic}
\usepackage{algorithm}
\usepackage{array}
\usepackage[caption=false,font=normalsize,labelfont=sf,textfont=sf]{subfig}
\usepackage{textcomp}
\usepackage{stfloats}
\usepackage{url}
\usepackage{verbatim}
\usepackage{graphicx}
\usepackage{cite}
\usepackage{booktabs}

\title{CLARA: Multilingual Contrastive Learning for Audio Representation Acquisition}

\author{
\IEEEauthorblockN{Kari A Noriy\IEEEauthorrefmark{1}\IEEEauthorrefmark{3}, 
Xiaosong Yang\IEEEauthorrefmark{1}\IEEEauthorrefmark{3}, 
Marcin Budka\IEEEauthorrefmark{2} and
Jian Jun Zhang\IEEEauthorrefmark{1}\IEEEauthorrefmark{3}}\\
\IEEEauthorblockA{\IEEEauthorrefmark{1}Centre for Digital Entertainment, \IEEEauthorrefmark{2}Computer Science and Informatics, \IEEEauthorrefmark{3}National Centre for Computer Animation\\
Bournemouth University\\
Email: \{knoriy, xyang, mbudka, jzhang\}@bournemouth.ac.uk}
}

\begin{document}

\maketitle

\begin{abstract}

In multilingual speech processing, accurately understanding and interpreting emotions is pivotal yet challenging due to the scarcity of labelled datasets. The recent strides in contrastive learning have catalysed self-supervised methodologies, enabling the harnessing of unlabelled data. In light of these advancements, we introduce CLARA, a groundbreaking multilingual framework, as our solution to diminish labelled data dependence and escalate generalisation across a diverse array of languages and conditions. CLARA excels in guiding models to embody shared representations across languages, seamlessly facilitating the cross-lingual transfer of speech and emotional understanding, even amidst scenarios with limited target language data. A cornerstone of our approach is proficiently capturing emotional subtleties within speech, transcending the challenges posed by the subjective nature of perceptual assessments. Embarking into self-supervised learning with a rich multilingual data corpus, we aim to delineate speech representations imbued with emotive dimensions, unlocking new potentials in emotion-aware multilingual speech processing.

Employing a substantial corpus of multilingual audio data, our methodology leverages data augmentation techniques to broaden the dataset spectrum, incorporating visual understanding via textual embedding, augmentation of language data from high-resource data sources to low-resource languages and model CLARA to learn these representations across domains.

Rigorous experimentation demonstrates our model's superior performance across various tasks, such as emotion recognition, multilingual language comprehension, audio classification, and retrieval benchmarks, especially in zero-shot and few-shot scenarios. Our model presents a compelling approach to obtaining shared and adaptable speech representations across languages and acoustic conditions while encoding latent emotional aspects. Additionally, we showcase the model's capability to adapt to low-resource languages, marking a significant stride in multilingual speech representation learning.
\end{abstract}

\noindent\textbf{Index Terms}: Contrastive Learning, multilingual speech, computational para-linguistics, text-to-audio retrieval, audio-to-text retrieval, audio classification.

\section{Introduction}

The domain of multilingual speech processing, particularly in natural and emotive speech representations, grapples with substantial challenges stemming from the requirement of extensively annotated datasets across a vast linguistic spectrum. Supervised learning methods that rely on large labelled corpora have achieved high performance in high-resource languages like English. Still, their effectiveness does not transfer well to low-resource languages where transcribed speech and emotionally labelled data in a myriad of environmental conditions are scarce. This data dependence limits the development of speech technologies for global users.

Recent research has explored semi-supervised and self-supervised approaches to learning from unlabelled data by leveraging data abundance to compensate for the lack of annotations. Contrastive learning has emerged as a promising technique for self-supervised representation learning in domains like computer vision and natural language processing. By training models to distinguish between similar and dissimilar sample pairs. The contrastive objective enables models to extract meaningful representations from unlabelled data.

Several studies have applied speech representations learning from audio. For instance, AudioCLIP \cite{paper:audioclip} demonstrates that contrastive losses can align environmental sounds and visual representations in a shared space. Still, this approach learns a shared representation of labelled data. COLA \cite{paper:coca} introduces a framework tailored for self-supervised speech representation learning through sample pairs drawn from diverse unlabelled audio, but this approach does not consider the lexical content and solely focuses on learning representations from a single modality (waveform). Furthermore, existing contrastive speech models are trained on single languages and need to fully leverage multilingual data.

Learning generalised multilingual natural language representations is crucial for advancing speech processing across low-resource languages. Hence, we advocate a multifaceted paradigm that encapsulates natural language text descriptors and the emotive states of speakers, aspiring to delve beyond mere lexical interpretation to a comprehensive understanding of the speakers.
We incorporate image-based embedding into our natural language text descriptions to further enhance this aspect of our work. This approach allows us to continue building on high-quality data such as EmoV-DB \cite{dataset:EmoV-DB}, where a video recording of the speaker is not available, reducing model complexity and aiding in faster learning.

Multilingual training exposes models to greater diversity in speakers, accents, acoustics, phonetics, and languages. This facilitates learning robust and transferable representations that perform well even for limited target language data. While supervised multilingual models require abundant labelled data, our semi-supervised approach provides an avenue to exploit the abundance of unlabelled multilingual speech.

Motivated by these challenges, we introduce an innovative multilingual framework geared towards assimilating universal speech and emotional representations, employing a rich tapestry of speech data across myriad languages. Our approach learns a shared representation space aligned across languages to enable positive transfer to low-resource languages. We train on an ensemble of unlabelled multilingual speech and video data. We also introduce novel data augmentation techniques that leverage visual cues without dependence on visual information and multilingual natural language captions, unlike similar models, further expanding the diversity of training samples.

This paper makes the following contributions: 1) A multilingual approach trained to extract a joint representation space from speech across languages while incorporating vision, textual descriptions and transcriptions. 2) Innovative data augmentation techniques that fuse visual and multilingual textual data, significantly amplifying the richness of both labelled and unlabelled datasets. 3) Through rigorous experiments, which unveil a state-of-the-art performance across various downstream speech processing tasks within zero-shot and few-shot frameworks, underlining the substantial advancements our framework brings forth in tackling the identified challenges.
These contributions provide an effective approach to learning joint speech and emotion representations that generalise across languages, accents, and acoustic conditions.

The subsequent sections will elaborate on our proposed approach, model training methodology, and present experimental results. We conclude by discussing key takeaways, limitations, and promising future directions.

\section{Related work}

Contrastive learning has become an effective technique for unsupervised and weakly supervised representation learning across modalities, including vision, language, and audio. Models trained with contrastive objectives learn robust features from unlabelled data by distinguishing between similar and dissimilar sample pairs.

Several pioneering works have successfully applied contrastive learning in computer vision. SimCLR \cite{paper:simCLR, paper:simCLRv2} introduced a simple framework for image representation learning by maximising agreement between differently augmented views of the same image and minimising agreement between views from different images. MoCo \cite{paper:moco} proposed a momentum contrast approach using a dynamic dictionary to achieve state-of-the-art image classification.

CLIP \cite{paper:clip} extends SimCLR by introducing a hybrid loss function that combines cross-entropy and contrastive loss. This approach enables CLIP to learn representations that are capable of language understanding and image recognition tasks, and it has been applied successfully in various applications such as image generation \cite{paper:dalle1, paper:dalle2, paper:stablediffusion} and machine translation \cite{paper:clipCap}. 

More recently, contrastive learning has been adapted to the audio domain. AudioCLIP \cite{paper:audioclip} builds upon the success of CLIP \cite{paper:clip} by extending its domain to audio and learning cross-modal representations of audio and visual data through a contrastive loss function. Similarly, COLA \cite{paper:COLA} introduced a self-supervised objective tailored to learn from diverse unlabelled audio data. However, existing contrastive models have focused on single languages.

Our work builds on these prior works by proposing a novel multilingual contrastive learning approach to acquire shared speech representations across languages. We leverage the abundance of diverse multilingual speech and automatic transcriptions, in contrast to supervised methods reliant on scarce labelled corpora. Our data augmentation techniques incorporating visual and textual information further expand training samples for robust representation learning.

In summary, our framework aims to capitalise on the successes of contrastive learning in other domains by tailoring it to the multilingual speech setting through innovations like cross-lingual pretraining strategies and multimodal data augmentations. This helps overcome the limitations of supervised-only multilingual models and single-language contrastive approaches for speech and enables several downstream tasks, such as emotion detection, speech-to-text, and expressive speech synthesis and allows users of such model to guide the model via natural language text prompts such as "A woman speaking is a joyful mood".

\section{Corpus} \label{sec:corpus}
Contrastive models benefit from large datasets. Due to the semi-supervised nature of these models, a considerable amount of analogous and heterogeneous sample pairs is essential to achieve meaningful representation. In contrast, when data is limited, the model may require a more in-depth comprehension of the underlying data structure to ensure accurate predictions. Furthermore, contrastive learning often involves training the model with a vast number of negative samples (i.e., pairs of dissimilar data points), which can be a resource-intensive process. Therefore, a larger dataset can expedite the training process by providing additional negative samples for the model to learn from.

With this objective in mind, we compiled a collection of openly available speech, music, and sound effects datasets to establish an ensemble dataset. Furthermore, we have employed various augmentation techniques to expand the dataset's magnitude and incorporate speech metadata in a natural manner, including emotional attributes, which we explore in detail in this section.

\subsection{Datasets}
In collaboration with Laion, we compile publicly accessible data sets, incorporating a fusion of meticulously gathered web-scraped data and openly available resources. This amalgamated data reservoir consists of pairs of audio and text, encompassing both naturally occurring linguistic interactions and artificially generated language constructs. The latter category encompasses supplementary metadata attributes, including linguistic origin, titular designations, transcriptions, and audio-visual content manifestations such as animal vocalisations or news presenter narratives. 

The collection surpasses 16,000 hours, with approximately 12 million samples attributed to male speakers. In comparison, approximately 3 million samples pertain to female speakers, and approximately 5 million samples remain unclassified regarding gender categorisation. Moreover, around 6,000 hours within this compilation are comprised of environmental sounds. Refer to Appendix \ref{appendix:Corpus} for a comprehensive elaboration on these constituent datasets.

Recent models such as CLIP \cite{paper:clip} and SimCLR v1/v2 \cite{paper:simCLR, paper:simCLRv2} have demonstrated the potential of large-scale models to comprehend contextual representation. Learning from a massive corpus offers several advantages over alternative techniques. For instance, loosely supervised natural language data obtained from the web is more accessible to scale than conventionally labelled data. Models can passively acquire knowledge from large amounts of loosely labelled data from the web. As demonstrated by CLIP \cite{paper:clip}, these models do not solely learn a representation but also establish connections between the representation and language, permitting the flexible zero-shot transfer.

\subsubsection{Data Preparation}
\begin{figure*}[!t]
    \centering
    \subfloat[]{\includegraphics[width=0.4\textwidth]{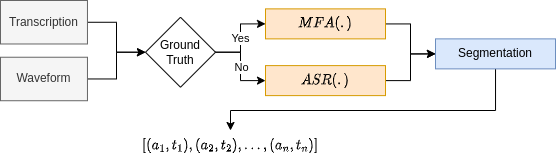}%
        \label{fig:dp_speech}}
    \hfil
    \subfloat[]{\includegraphics[width=0.4\textwidth]{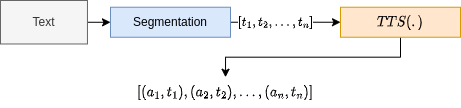 }%
        \label{fig:dp_syn_speech}}
    \caption{The above outlines the two frameworks for generating and augmenting existing data for speech. \ref{fig:dp_speech} outlines Speech-Aware Segmentation: Let \(MFA\) be a forced aligner generating a time-aligned transcription and \(ASR\) be an automatic speech recognition model. Let \(a_n, t_n\) be a meaningful truncated waveform and transcription. If no ground truth transcript is provided, \(ASR\) is used. \ref{fig:dp_syn_speech} outlines a string tokeniser and a Text-To-Speech model for generating speech, Let \(segmentation\) be a rule-based string tokeniser and \(TTS(.)\) be a neural speech synthesiser generating \(a_n\) audio and it corresponding transcript \(t_n\).}
    \label{fig:dp}
\end{figure*}

A standardised format is employed to systematically organize audio-text pairs. Audio samples undergo FLAC conversion and resampling to 48kHz for lossless compression and high-quality distribution, aligning with the EMNS dataset guidelines \cite{dataset:emns}. Metadata, containing descriptions like "A \(gender\) saying \(transcription\) in a \(emotion\) voice," is stored in JSON format, where \(gender\), \(transcription\) and \(emotion\) are datapoint from a given dataset used to form a natural language description. .tar files are used for scalable streaming. For extended audio exceeding 10 seconds, two truncation strategies are implemented. The first employs Montreal Forces Aligner (MFA) \cite{paper:MFA} for precise alignment of speech recordings and transcripts, while the second uses Whisper \cite{paper:whisper}, a transformer-based ASR system, to transcribe and truncate lengthy samples. The data pipeline segmentation is visually depicted in Figure \ref{fig:dp_speech}.


\begin{figure}[t]
    \centering
    \includegraphics[width=0.3\textwidth]{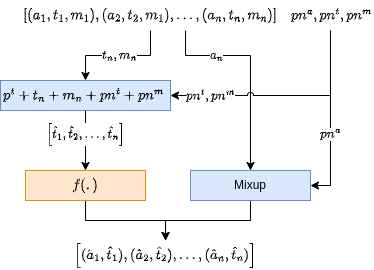 }
    \caption{The proposed framework provides a straightforward approach for augmenting natural language descriptions and incorporating environmental noise layering. Let \(a_n, t_n\) be given audio and text pairs, let \(p^t\) be a prefix string to prepend, and \(m_n\) and \(pn^n\) are additional meta information, such as audio quality, gender and emotion. Let \(f(.)\) be a schema combining labels to natural language description. Let \(pn^a\) (pub noise) be a series of environmental sounds and \(pn^t\) labels. \(\hat{a}_n, \hat{t}_n\) be the augmented audio and natural language description.}
    \label{fig:dp_mixup}
\end{figure}

\subsubsection{Multimodal Fusion and Text Generation} \label{sec:data aug}
Chen et al. \cite{paper:simCLR} demonstrated the significant impact of data augmentation on improving the performance of contrastive learning within the computer vision field. Other studies have also indicated that data augmentation can enhance the generalisation of supervised neural networks and combat overfitting \cite{paper:wei_comparison_2020, paper:zhang_mixup_2018}. Augmentation techniques increase the diversity of acoustic environments and recording conditions, acquiring robust and invariant features and ultimately regularising the model. Accordingly, we employ various augmentation techniques to diversify speech data for our model.

Unlike previous methods where class labels or simple templates are used. Our augmentation strategy encompasses a comprehensive set of techniques to broaden the diversity of our audio data. These techniques include both traditional audio augmentation methods and neural approaches (Figure \ref{fig:dp_mixup}). In traditional audio augmentation, we apply various transformations such as adding reverb, clipping, masking, pitch modulation, and introducing environmental sounds to simulate different acoustic environments or distorted audio. These transformations are layered on top of the original audio signals to create augmented versions.

\begin{equation} \label{eq:dp_mixup}
    \begin{aligned}
        \hat{a} =  \lambda a_i + (1-\lambda) a_j \\
        \hat{t} =  f(p^t + t_i + t_j)
    \end{aligned}
\end{equation}

For example, we generate augmented audio samples by combining two raw audio signals, \(a_i\) and \(a_j\), using a scale factor \(\lambda\) that ranges between 0.01 and 0.8 used to control the contribution of environmental sounds (pub noise), where \(a_i\) is a speech segment and \(a_j\) the background sounds to be added. The resulting augmented audio \(\hat{a}\) is a linear combination of the two signals. Similarly, the corresponding text labels \(t_i\) and \(t_j\) are combined using a function \(f(.)\) that employs concatenative and neural-based approaches to generate natural language descriptions. This augmentation pipeline is visually depicted in Figure \ref{fig:dp_mixup}.

To overcome the lack of large multilingual datasets for sound event classification. We made use of pre-trained multilingual language models to translate labels into different languages and overlay them on mono-lingual data, synthesising multilingual datasets. This efficient data augmentation approach leverages language models to avoid costly manual collection, rapidly creating diverse training corpora covering many languages to enable multilingual fusion for low-resource language.

To provide a concrete example, consider the following concatenated text: "A person saying \textit{It's raining outside}, background \textit{wind noise}", where \(t_i\) and \(t_j\) are `It's raining outside' and `wind noise' respectively and \(p^t\) is a given template `A person saying \{\}, background \{\}'. To obtain meaningful text descriptions from this concatenated sentence, we leverage the Language Models such as Open Assistant \cite{site:openAssistant} to generate natural language descriptions by processing the concatenated text, incorporating context, and producing coherent and human-like outputs. In our example, the language model generates a description like "Amidst wind noise in the background, an individual declares, 'It's raining outside'". We further use CoCa to generate visual captions for samples with both audio and visual elements, employing Structural Similarity Index (SSIM) to identify significant changes in visual frames and generate visual captions accordingly. These visual captions are combined with audio descriptions using a language model to create a unified representation of multimedia content, enhancing caption generation.

In Appendix \ref{appendix:Augmentation}, we elaborate on our augmentation strategy, detailing additional techniques and their specific implementation.

\section{Method}
\begin{figure*}[!t]
    \centering
    \subfloat[]{\includegraphics[width=.31\textwidth]{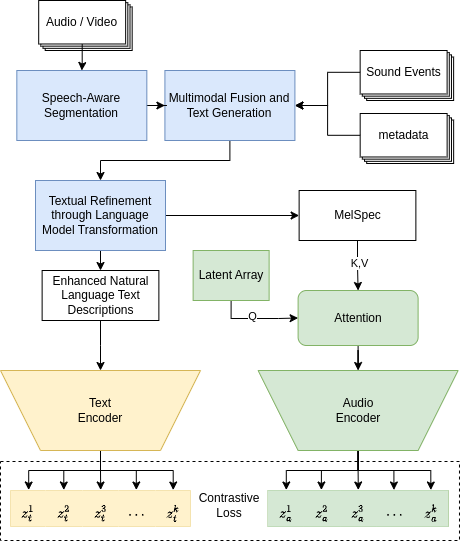}
        \label{fig:CLARA_model}}
    \hfil
    \subfloat[]{\includegraphics[width=0.39\textwidth]{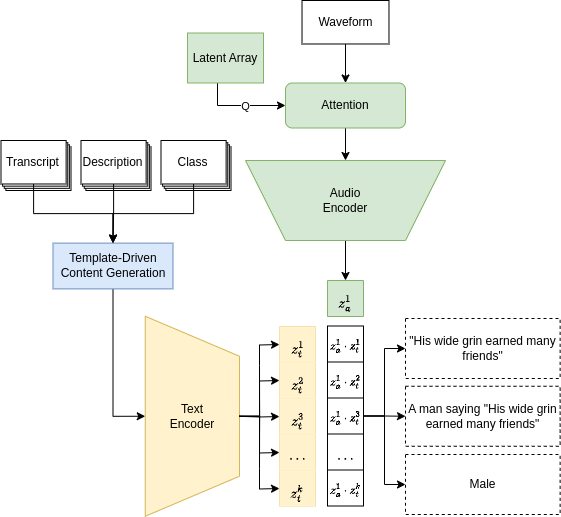}
         \label{fig:CLARA_tasks}}

    \caption{The above figures present an overview of our proposed CLARA approach, highlighting its structural architecture and key role in enhancing audio-text feature learning and related tasks. \ref{fig:CLARA_model} describes our contrastive framework designed for learning audio and text feature representations. This framework employs Speech-Aware segmentation combined with scene change detection, multimodal fusion, and text description refinement. \ref{fig:CLARA_tasks} illustrates the pretrained encoders within the model that map both audio inputs and corresponding text data into a shared representation space. A Perceiver \cite{paper:perceiver} head is used to handle a vast number of input data points, such as those from long-form audio. This unified space enables accurate classification and retrieval of class labels, transcriptions, or text descriptions corresponding to the input audio.}
    \label{fig:CLARA_framework}
\end{figure*}

We train two encoder networks (depicted in Figure \ref{fig:CLARA_model}) to learn mixed input representations. Our training approach employs a contrastive loss function to optimise the similarity between matching pairs and minimise the similarity between dissimilar pairs.

\begin{equation}
    \begin{aligned}
        z_a =  g(f_a(A)) \\
        z_t =  g(f_t(T)) \\
    \end{aligned} \label{eq:CLARA}
\end{equation}

The method employed is demonstrated through equation \ref{eq:CLARA} and visualised in figure \ref{fig:CLARA_model}. The Audio Encoder \(f_a\) processes log mel spectrogram \(A\) to generate corresponding hidden representation \(z_a\); our audio encoder utilises latent vectors, providing an efficient acoustic encoder with a fixed number of latent variables. This approach allows CLARA to process inputs like audio without linearly scaling the number of parameters with the input size. The Text Encoder \(f_t\) produces the hidden representation \(z_t\) by processing text samples \(T\) using a sinusoidal position embedding. Finally, both encoders are connected to a residual projection head \(g(.)\) that maps the encoder representation to the contrastive loss space.

The encoders incorporate a transformer architecture with a 1024 embedding. The text encoder is a transformer utilising Flash Attention \cite{paper:flashattention}, our audio encoder utilises the Perceiver head \cite{paper:perceiver} for handling large input data points, and we modify the encoder by adding a convolution head and GELU activation to the key and values. The latent vector provides an efficient acoustic encoder with a fixed number of latent variables, allowing it to process inputs like audio without scaling the number of parameters linearly with the input size, while flash attention increases modelling capacity and training efficiency. Both encoders employ Cosine positional encoding to integrate positional information of tokens within the input sequence effectively.

\subsection{Loss}
We adopt a refined CLIP loss function inspired by \cite{paper:clip, paper:simCLR} to map audio and text inputs towards a shared latent space. This alteration offers distinct advantages over the conventional contrastive loss framework, enabling the concurrent attainment of multi-modal alignments and the extraction of semantic nuances from both modalities. The evolved representation thus orchestrates the cohesive clustering of audio and text based on semantic similarity within the latent space, accounting for both magnitude and direction.

In this adapted context, the loss function is expressed as:

\begin{multline}
     \mathcal{L} = -\frac{1}{N} \sum_{k=1}^N \left[ log \frac{exp((z^k_a \cdot z^k_t)\tau_a)}{\sum^N_{n=1} exp((z^k_a \cdot z^n_t)\tau_a)} + \right.\\
     \left. log \frac{exp((z^k_t \cdot z^k_a)\tau_t)}{\sum^N_{n=1} exp((z^k_t \cdot z^n_a)\tau_t)}\right]
\end{multline}

This equation introduces the learnable temperature parameters \(\tau_a\) and \(\tau_t\) for audio and text, respectively. The terms \(z_a^k\) and \(z_t^k\) denote the \(k\)-th pair of audio and text features, while \(N\) represents the minibatch size. Through this augmented loss formulation, we seek to enhance the discriminative potential of the shared latent space, leveraging individualised temperature controls for each modality to capture intricate relationships more effectively.



\section{Experiments}
We conducted extensive experiments to evaluate our proposed multilingual contrastive learning framework. Models were assessed on a diverse set of speech processing benchmarks under zero-shot and few-shot conditions. 

To evaluate the multilingual performance, we use Keyword spotting (KWS). KWS demonstrates language-agnostic representation learning, capturing the underlying acoustic patterns and demonstrating the ability to adapt to low-resource languages. In addition, we perform zero-shot (ZS) classification and retrieval, which evaluates our model on tasks not directly trained on, such as emotion detection and sound event classification. Zero-shot allows us to gauge our model's ability to generalise its learned knowledge to tasks it hasn't been exposed to during training and its ability to transfer knowledge to downstream tasks. To further evaluate the representations learned by our model, we perform a Linear Probe (LP). Liner probe is the process of training a simple linear classifier on top of our pre-trained network, which allows us to evaluate the quality of the learned representations by assessing how well they can be used for a specific downstream task without any further finetuning or adaptation of our model. Finally, we look into information retrieval; retrieval demonstrates understanding and incorporation of contextual information from the input queries.


We juxtapose our model against state-of-the-art contrastive models documented in the existing literature, encompassing zero shot (ZS) and supervised/finetuned approaches (Best), to determine its performance under various scenarios. Furthermore, we compare with contemporary models trained for a single task (SOTA).

\begin{table*}
  \caption{In emotion and audio classification, our model surpasses prior approaches through zero-shot (ZS) and fine-tuned (Best) assessment. It achieves leading zero-shot results on CREMA-D, RAVDESS, Audioset, US8K, and FSD50K. Fine-tuning enhances performance, particularly on CREMA-D, RAVDESS, US8K and FSD50K, surpassing previous models. These outcomes highlight our model's learned generalised representations, obtained through multimodal pretraining, which excel in zero-shot and fine-tuned evaluations. SOTA are model trained solely for a single task.}
  \label{tab:Eval}
  \centering
  {\fontsize{7}{9}\selectfont
  \begin{tabular}{ l|cc|cccc }
    \toprule
                             & \multicolumn{2}{c}{Emotion} & \multicolumn{3}{c}{Sounds} \\ 
                                \cmidrule(lr){2-3}  \cmidrule(lr){4-6}
        Model                &  \shortstack{CREMA-D\\(Acc)} & \shortstack{RAVDESS\\(Acc)} & \shortstack{Audioset\\(mAP@1)} & \shortstack{US8K\\(Acc)} & \shortstack{FSD50K\\(mAP@1)}\\
    \midrule
        CLAP (ZS) \cite{paper:ms_clap}        & 17.84    & 15.99     & 05.80     & 73.24    & 30.24 \\
        CLAP (Best) \cite{paper:ms_clap}      & 68.34    & 64.36     & -         & 87.96    & 58.59 \\
        AudioCLIP (ZS) \cite{paper:audioclip} & -        & -         & -         & 65.30    & -     \\
        AudioCLIP \cite{paper:audioclip}      & -        & -         & 28.36     & 89.49    & -     \\
        Wav2CLIP (ZS) \cite{paper:wav2clip}   & -        & -         & -         & 40.44    & 03.02 \\
        Wav2CLIP \cite{paper:wav2clip}        & -        & -         & -         & 81.01    & 43.08 \\
    \midrule
        ours (ZS)                             & \textbf{54.87}    & \textbf{26.33}     & \textbf{13.49}     & \textbf{87.84}    & \textbf{32.90} \\
        ours (Best)                           & \textbf{87.14}    & \textbf{85.01}     & 22.21              & \textbf{89.20}    & \textbf{61.35} \\
    \midrule
        SOTA       & 70.95 \cite{paper:larc}    & 84.10 \cite{paper:VQ-MAE-S-12}     & 47.10 \cite{koutini_efficient_2022}     & 90.00 \cite{paper:eat} & 69.70 \cite{paper:onepeace}\\
    \bottomrule
  \end{tabular}
  }
\end{table*}

\subsection{PreProcessing}
The audio data preprocessing pipeline converts raw audio waveforms to log Mel spectrogram representations. The spectrograms are computed using a sampling rate of 48 kHz, 80 Mel filterbanks spanning the frequency range 0-8 kHz, a hop length of 512 samples between successive frames, and a Hann window size of 1024 samples. Randomly selected audio clips are truncated or zero-padded to match the duration of the longest sample within each training batch. This standardised spectrogram representation allows the model to learn robust audio features for the downstream task.

\subsection{Multilingual Keyword Spotting}

An analysis of the empirical results attained by our proposed CLARA model on the Multilingual Spoken Words Corpus (See figure \ref{fig:KWS}), as detailed further in Appendix \ref{appendix:KWS}, demonstrates that while sufficient amounts of labelled training data are indispensable for achieving superior keyword spotting performance in a given target language, synergistic transfer of linguistic knowledge from closely related or cognate languages with abundant annotated audio can also confer substantial accuracy benefits. This is evidenced by CLARA's ability to generalise well even when trained on reasonably small sample sizes for low-resource languages like Odia, Lithuanian, and Guarani, attaining 61.5\%, 93.7\%, and 88.8\% accuracy, respectively, with just 1-6 minutes of audio data or between 100-300 samples. Our model capitalises on overlapping linguistic typologies and shared roots with other languages, transferring representations of phonology, morphology, and lexicon. Cross-lingual transferability appears less pronounced for isolated or linguistically distinct languages, underscoring the importance of language family relations and genealogical proximity for enabling efficient transfer learning in low-resource multilingual keyword spotting tasks. Nonetheless, these results highlight CLARA's promising capabilities for generalised multilingual keyword spotting across diverse languages.

\begin{figure}
    \centering
    \includegraphics[width=0.43\textwidth]{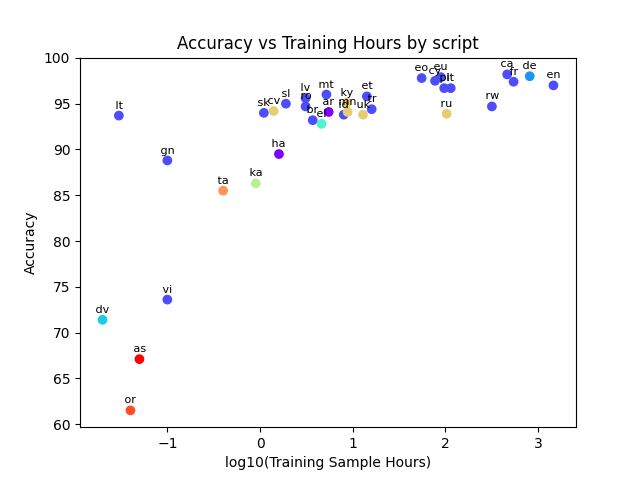 }
    \caption{The relationship between the quantity of training data and keyword spotting accuracy across 36 languages from the Multilingual Spoken Words Corpus is illuminated, with each data point representing an individual language's performance. High-resource languages such as English and French, with abundant training data, exhibit elevated accuracy, while low-resource kindred languages still achieve proficient results through presumed transfer of representations within language families. However, more isolated languages like Vietnamese necessitate greater data to learn robust representations, as evidenced by poorer accuracy despite similar training hours to Guarani. Overall, the results highlight the influence of training data size on model performance and demonstrate the potential for cross-lingual transfer of learned representations to enable generalised accuracy for low-resource languages, provided sufficient data exists across related languages.}
    \label{fig:KWS}
\end{figure}

\subsection{Zero-shot}
Zero-Shot explores the generalisation of CLARA to unseen samples. By training a model using contrastive learning, we facilitate mapping these attribute representations into a shared semantic space, enabling the model to reason about unseen classes based on their relationships with seen classes. This experiment investigates the effectiveness of zero-shot learning in recognising and classifying previously unseen classes, showcasing the potential for knowledge transfer and generalisation. Our findings shed light on the capabilities and limitations of zero-shot learning in real-world scenarios, offering insights into its applicability and performance.

\subsubsection{Setup}
We construct a natural language description from the given class label. Different templates were used for different tasks. For instance, we used 'The sound of \{label\}' for sound classification and 'A person talking in a \{label\} voice' for emotion classification.


\subsubsection{Results}
In zero-shot classification tasks (Table \ref{tab:Eval}), our model achieves superior performance compared to existing contrastive models on emotion recognition benchmarks CREMA-D and RAVDESS, with improvements of up to 300\% on CREMA-D. Specifically, our zero-shot accuracy of 54.87\% on CREMA-D significantly exceeds the 17.84\% achieved by CLAP, representing a 3x increase. Similarly, on RAVDESS, our 26.33\% zero-shot accuracy exceeds CLAP's performance by a factor of 1.6x.


Additionally, our model establishes new state-of-the-art zero-shot results on the sound event classification datasets Audioset, US8K, and FSD50K. On Audioset, CLARA attains 13.49\% zero-shot accuracy, surpassing CLAP (5.80\%). On US8K, we achieves 87.84\% accuracy, outperforming CLAP (73.24\%), AudioCLIP (65.30\%), and Wav2CLIP (40.44\%). Finally, on FSD50K, CLARA's Mean Average Precision (mAP) of 32.90\% exceeds CLAP (30.24\%) and Wav2CLIP (3.02\%), representing a 2.66\% increase over previous methods.

\subsubsection{Zero-Shot Templates} 

Natural language templates are crucial for zero-shot learning, as their wording and framing fundamentally affect generalisation. More unambiguous templates improve accuracy on unseen classes, while conceptual prompts degrade performance. We use simple descriptive prompts for each target class. For instance, 'A person talking in a \{label\} voice' for emotion detection. Table \ref{tab:zs_templates} shows templates for different tasks. The choice of template impacts metrics like accuracy, as abstract prompts require more reasoning and cause declines as shown in table \ref{tab:zs_templates_eval} on sound event classification. Careful design is needed to maximise inclusion and avoid biases. Template engineering is vital for zero-shot learning, balancing clarity, specificity and inclusivity.

\begin{table}
    \caption{Natural language templates used to generate descriptive phrases for various zero-shot classification tasks. \{label\} is replaced by the target class name during prompt generation.}
    \label{tab:zs_templates}
    \centering
    \begin{tabular}{ll}
        \toprule
            \textbf{Task}               & \textbf{Template}                         \\
        \midrule
            Age                         & 'A \{label\} year old'                      \\
            Gender                      & 'A \{label\}'                               \\
            Emotion                     & 'A person talking in a \{label\} voice'     \\
            Sound                       & 'The sound of \{label\}'                    \\
            Language                    & '\{label\} speaker'                         \\
        \bottomrule
    \end{tabular}
\end{table}

\begin{table}
    \caption{Accuracy on zero-shot sound classification with different natural language prompt templates. More concrete templates like "{label}" and "The sound of a {label}" achieve higher accuracy. More conceptual prompts like "The {label} emits a distinctive sound" reduce accuracy, requiring increased reasoning. The phrasing and specificity of templates substantially impacts zero-shot generalization, underscoring the importance of careful template design}
    \label{tab:zs_templates_eval}
    \centering
    \begin{tabular}{lc}
        \toprule
            \textbf{Template}                         & \textbf{Acc}             \\
        \midrule
            '\{label\}'                               & 89.09                         \\
            'A \{label\}'                             & 89.09                         \\
            '\{label\} sounds'                        & 89.20                         \\
            'A \{label\} can be heard'                & 66.25                         \\
            'The sound of a \{label\}'                & 89.32                         \\
            'The \{label\} emits a distinctive sound' & 55.00                         \\
        \bottomrule
    \end{tabular}
\end{table}

\subsection{Linear probe (LP)}

Following established best practices, we conducted a linear probing analysis to evaluate the pre-trained model's ability to learn representations that transfer to novel downstream tasks. Specifically, we appended a randomly initialised linear evaluation head on top of our pre-trained CLARA model. This linear probe comprises three fully connected layers with ReLU activations and dropout regularisation. LP allows us to analyse the transferability of the base model's learned representations by training the linear probe model to minimise cross-entropy loss datasets for downstream tasks.

\subsubsection{Setup}
We froze the parameters of the pre-trained base model and attached a simple classifier to the tail of our model. The classification tail is optimised end-to-end using the AdamW optimiser with a small learning rate of \(1e-4\). The training (Table \ref{tab:exp_emotion_LP}) was limited to a maximum of 20 epochs except for \textit{Best} to prevent overfitting, all models utilise validation loss for early stopping. This framework allows assessing the generalisable feature learning abilities of the base model without degrading performance on its original pretraining task.

\subsubsection{Results}

Our linear probe (table \ref{tab:exp_emotion_LP}) experiments demonstrate the efficacy of our model's learned representations for downstream emotion and gender classification tasks without explicit training on these targets. The linear probe (LP) consistently outperforms the base zero-shot (ZS) model on both EmoDB and CREMA-D benchmarks. For emotion classification, the LP achieves 90.93\% top-1 recall on EmovDB versus just 42.13\% for the ZS baseline and 68.46\% versus 42.00\% on CREMA-D. Despite our model not being specifically trained to classify the emotions, we observe significant gains in recall scores, demonstrating that our base model has acquired generalisable internal representations containing substantial information about emotional content. Furthermore, the LP reaches performance comparable to state-of-the-art models trained solely for emotion classification, substantiating the versatility of our general-purpose learned features. on RAVDESS we outperform VQ-MAE-S-12 \cite{paper:VQ-MAE-S-12} by 2.8\% absolute and achieve comparable results on CREMA-D to LeRaC \cite{paper:LeRaC} (70.95\%). Our model exhibits an aptitude for intrinsic emotional modelling, as evidenced by the linear probe's strong exploitability of these internal representations for emotion and gender predictions without task-specific finetuning.

\begin{table*}
  \caption{Our model's versatile pretraining on EmoV-DB, CREMA-D, and RAVDESS datasets shows superior zero-shot retrieval and linear probing performance in emotion and gender classification. The model's strong generalisation, learned during pretraining, leads to competitive few-shot transfer results compared to previous state-of-the-art models focused on individual tasks.}
  \label{tab:exp_emotion_LP}
  \centering
  {\fontsize{7}{9}\selectfont
      \begin{tabular}{ c|cc|ccc|c|c }
        \toprule
                            & \multicolumn{2}{c}{Emotion (ZS)} & \multicolumn{3}{c}{Emotion (LP)} & \multicolumn{1}{c}{Gender (ZS)} & \multicolumn{1}{c}{Gender (LP)} \\
                                                \cmidrule(lr){2-3}      \cmidrule(lr){4-6} \cmidrule(lr){7-7}      \cmidrule(lr){8-8}
                            Training Dataset    & R@1   & R@3   & R@1   & R@3   & Acc   & Acc   & Acc \\
        \midrule
                            EmovDB              & 42.13 & 83.22 & 90.93 & 99.59 & 90.12 & 58.35 & 99.05 \\
                            CREMA-D             & 42.00 & 79.63 & 68.46 & 85.20 & 69.53 & 61.94 & 91.61 \\
                            RAVDESS             & 58.12 & 90.54 & 88.44 & 99.24 & 86.92 & 94.62 & 94.65 \\
        \bottomrule
      \end{tabular}
  }
\end{table*}

\subsection{Retrieval}
The retrieval process of both audio and text is accomplished through a series of steps involving the use of the corresponding audio encoder (\(f_a(.)\)) and text encoder (\(f_t(.)\)) for capturing and representing the audio and text samples, respectively. These encoders transform the audio and text data into high-dimensional feature vectors (embeddings) that encode the relevant information needed for effective retrieval. The embeddings are projected into a shared contrastive space via the projection function \(g(.)\), which enables the comparison of the audio and text embeddings. Cosine similarity is then calculated between the projected audio and text embeddings to determine the degree of semantic alignment. This pipeline allows retrieving the most relevant text passage for a query audio clip and vice versa, as illustrated in Figure \ref{fig:CLARA_model}.

\begin{table*}
  \caption{Performance of audio-to-text and text-to-audio retrieval on the AudioCaps dataset. Our proposed model achieves state-of-the-art results, outperforming prior models CLAP and ML-ATCMR on metrics including R@1, R@5, and R@10 for both retrieval directions.}
  \label{tab:exp_sounds_retrieval}
  \centering
  {\fontsize{7}{9}\selectfont
      \begin{tabular}{ lc|ccc|ccc }
        \toprule
                                       && \multicolumn{3}{c}{Audio \(\rightarrow\) Text} & \multicolumn{3}{c}{Text \(\rightarrow\) Audio}\\
                                                            \cmidrule(lr){3-5}                 \cmidrule(lr){6-8}
            Model                           &\shortstack{Retrieval\\Dataset} & R@1   & R@5   & R@10  & R@1   & R@5   & R@10  \\
        \midrule
            CLAP \cite{paper:ms_clap_reteve}& AudioCaps                      & 30.81 & 41.91 & 73.18 & 34.69 & 70.22 & 82.00 \\
            ML-ATCMR \cite{paper:ml-atcmr}  & AudioCaps                      & 39.40 & 72.00 & 83.90 & 39.40 & 72.00 & 83.90 \\
            VAST \cite{paper:vast}          & AudioCaps                      & -     & -     & -     & \textbf{52.00} & 76.80 & 82.90 \\
        \midrule
            Ours                            & AudioCaps                      & \textbf{42.08} & \textbf{81.57} & \textbf{93.00} & 41.76 & \textbf{80.12} & \textbf{91.16} \\
        \bottomrule
      \end{tabular}
  }
\end{table*}

\begin{table*}
  \caption{Retrieval results on datasets containing paralinguistic features such as emotion and intensity using the EmoV-DB (EDB), CREMA-D (CD), and combined EDB+CD+RAV datasets for pertaining, where RAV is RAVDESS Corpus \cite{dataset:RAVDESS}.}
  \label{tab:exp_emo_retrieval}
  \centering
  {\fontsize{7}{9}\selectfont
      \begin{tabular}{ c|cccc|cccc }
        \toprule
                            & \multicolumn{4}{c}{Audio \(\rightarrow\) Text} & \multicolumn{4}{c}{Text \(\rightarrow\) Audio}\\
                                                    \cmidrule(lr){2-5}              \cmidrule(lr){6-9}
            Training Dataset    & mAP@1 & mAP@3 & mAP@5 & R@1    & mAP@1 & mAP@3 & mAP@5 & R@1   \\
        \midrule
            EmovDB              & 78.40 & 87.76 & 88.01 & 78.19  & 78.54 & 87.65 & 88.02 & 78.30      \\
            CREMA-D             & 92.71 & 96.24 & 96.24 & 90.97  & 87.85 & 93.46 & 93.37 & 86.11      \\
            EDB + CD + RAV      & 71.74 & 83.29 & 84.05 & 71.49  & 69.65 & 82.05 & 82.80 & 69.44      \\
        \bottomrule
      \end{tabular}
  }
\end{table*}

\subsubsection{Results}
Table \ref{tab:exp_sounds_retrieval} shows that our proposed model achieves state-of-the-art performance on both audio-to-text and text-to-audio retrieval benchmarks. In audio-to-text retrieval on the AudioCaps dataset, our model outperforms prior work, increasing R@1 by 11.27\% absolute over the current CLAP model and 2.68\% over ML-ATCMR. More gains are observed at R@5 and R@10 ranks, with our model demonstrating 39.66\% and 19.82\% improvements over CLAP, and 9.57\% and 9.10\% over ML-ATCMR, respectively. 

For text-to-audio retrieval, our model exhibits strong R@5 and R@10 results, surpassing existing models by 3.32\% and 8.26\% absolute. The considerable gains in lower-rank performance highlight our model's capabilities in recall-oriented retrieval, leveraging versatile relevance matching and broad conceptual knowledge. While precision at top-1 is lower, our model nonetheless produces nuanced and useful retrievals beyond the most obvious result. These comprehensive gains underscore the efficacy of our approach for both precision- and recall-oriented multi-modal retrieval.

\section{Conclusions}

This paper introduced CLARA, a novel framework for multilingual learning of speech and emotion representations. CLARA adeptly assimilates shared representations harmonised across diverse linguistic landscapes, fostering a transference of learned attributes to novel languages and tasks.

In our exploration of data augmentation, we capitalised on visual cues derived from video content to glean emotional and contextual insights. These insights were transformed into text format and injected into the natural language descriptors affiliated with audio data. This approach eliminates the need for creating new datasets where both audio and video must be present, allowing us also to use existing audio datasets where video data is unavailable. This further ensures flexibility and enables this work to expand into areas where video data cannot be captured.
Utilising advanced neural methodologies, we amplified textual meta-information and cues, aiming to enrich the training sample spectrum across a plethora of languages. This endeavour facilitated the generation of coherent natural language descriptions that encapsulate the contextual essence of the audio content. Our carefully designed experimental framework bore testimony to the robust performance of the model. This achievement indicates a significant stride towards enhancing state-of-the-art benchmarks in the domains of emotion recognition, audio classification, and retrieval tasks. The model showcased efficacy under zero-shot and few-shot scenarios, thereby underscoring its potential in navigating the complexities inherent in multilingual speech processing tasks.

The empirical outcomes accentuate the merits of our framework in procuring universally generalisable speech representations that thrive even amidst the scarcity of target language data. The versatility and adaptability of the model were further corroborated by the competitive performance of simple linear classifiers, which, when employed on frozen base features, required no task-specific fine-tuning. Some limitations were observed in extending representations to isolated languages. Future work should explore approaches tailored to the specific characteristics of individual languages, especially those that are isolated and consult with linguistic experts to better understand the nuances and intricacies of isolated languages to address these weaknesses. Nonetheless, our results highlight CLARA's promising capabilities for generalised learning of emotion and speech across diverse languages.

This study marks notable milestones, unveiling a model for speech and emotion understanding, data augmentation techniques incorporating visual, meta and audio data, and robust empirical validation of superior performance across various speech-processing tasks. It underscores the potential of this methodology in harnessing universal speech representations that transcend linguistic, accentual, and acoustic diversities, thus paving the way for groundbreaking advancements in the domain of multilingual speech and emotion processing.

\section{Acknowledgements}
This research was supported by The Engineering and Physical Sciences Research Council (EPSRC) by means of a grant awarded to Kari A Noriy. CDE2: EP/L016540/1 (for 2014/15 cohort and subsequent intakes), and we would like to thank the support of Stability AI for providing compute resources that were instrumental in conducting the research presented in this publication and the Laion community member who has helped collect and process the dataset used in training this model.

\bibliographystyle{IEEEtran}
\bibliography{mybib}

\include{appendix}

\end{document}

%% file: appendix.tex
\newpage
\appendix

\section{Corpus}\label{appendix:Corpus}

\subsection{Training Dataset} \label{appendix:Augmentation}

The training dataset consists of diverse publicly available audio-text pairs and sound event datasets from various sources. In total, the ensemble dataset consists of over 16,000 hours of audio matched with textural description, equating to approximately 21 Million audio text pairs. An essential data constituent comprises naturally occurring speech recordings extracted from datasets including Common Voice, Multilingual Spoken Words (MSWC) and LJSpeech. These datasets collectively provide thousands of hours of read speech originating from audiobook narrations, public addresses, and crowdsourced contributions in a multitude of languages. The inherent diversity in speakers, accents, ambient noise conditions, and discourse topics induces a rich representation of real-world speech.

Additionally, synthetic descriptions are obtained from audio captioning datasets such as Clotho, AudioCaps, and Audiocaps, comprising human-authored captions summarising the essence of brief audio snippets. These descriptive sentences provide semantic textual portrayals of sounds and acoustic events.

To further augment the training data, unlabeled audio segments are paired with generated captions and metadata by models such as CoCa and Whisper. CoCa is an image-text foundation model that produces natural language descriptions for visual scenes. Whisper is an automatic speech recognition system capable of transcribing speech. These models facilitate the synthesis of captions that expand the diversity of linguistic styles and speech contexts displayed in the data.

The extensive and multifaceted dataset is indispensable for training an efficacious contrastive speech representation model per our proposed approach. The corpus construction methodology aims to engender expansive coverage of linguistic phenomena through diverse speech recordings paired with descriptive captions.

\subsubsection{Demographic}

Table \ref{tab:datasets} provides an overview of the key details for each dataset utilised for training, including the type of audio content (such as speech, emotion, age, gender), total hours of audio, number of languages represented, and speaker gender metadata. The composite multilingual dataset enables robust representation learning by the model across 181 languages, various accents, age groups, genders, and acoustic environments. The diversity of speech, spoken words, emotions, ages, genders, and environmental sounds enhances the model's semantic parsing capabilities for both human vocals and ambient sounds. 

The top 5 languages in terms of audio sample size including train, validation and test splits are English (4,240 hours), Catalan (2,038 hours), Kinyarwanda (1,864 hours), German (1,835 hours) and French (1,434 hours). For English, the gender distribution is imbalanced, with 18,925 male and 27,828 female samples, along with 75,045 samples with unknown gender. Understanding such distributions across languages and demographics will inform active learning to improve model generalisation. Overall, the tabular overviews offer essential insights into sample diversity, gaps, and opportunities from a multilingual and gender representation perspective to guide the ongoing enhancement of the composite dataset. 

\begin{table*}
  \caption{CLARA training uses various audio datasets encompassing speech (SP), emotion (E), age (A), gender (G), spoken words (W), and environmental sounds (S). Details provided include total audio hours, languages per dataset, and speaker gender information, including male (M) and female (F) speakers, along with samples with unknown gender (U). These diverse, multilingual datasets empower CLARA to acquire robust audio representations spanning languages, accents, ages, genders, and environments.}
  \label{tab:datasets}
  \centering
  \begin{tabular}{lllllll}
    \toprule
    \textbf{Name}                                                   & \textbf{Content}    & \textbf{$\approx$ Hours}    & \textbf{Languages}  &\textbf{M} & \textbf{F} & \textbf{U}       \\
    \midrule
    Common Voice \cite{dataset:common_voice}                        & SP/G                & 3,098                       & 181                 & 472,021   & 185,930    & 306,779          \\
    MSWC \cite{dataset:mswc}                                        & W/G                 & 6,000                       & 50                  & 12,390,463& 3,060,697  & 5,455,351        \\
    YouTube Dataset                                                 & SP                  & 4000                        & 16                  & -         & -          & -                \\
    LibriSpeech \cite{dataset:librispeech}                          & SP                  & 1,000                       & 1                   & 1,283     & 1,201      & 0                \\
    LJSpeech \cite{dataset:librispeech}                             & SP                  & 24                          & 1                   & -         & 13,090     & -                \\
    CREMA-D \cite{dataset:CREMA-D}                                  & SP/E/G              & 5                           & 1                   & 3,930     & 3,512      & 0                \\
    EMNS \cite{dataset:emns}                                        & SP/E/A              & 2.5                         & 1                   & 0         & 1,205      & 0                \\
    EmoV DB \cite{dataset:EmoV-DB}                                  & SP/G                & 9.5                         & 1                   & 3,316     & 3,577      & 0                \\
    RAVDESS \cite{dataset:RAVDESS}                                  & SP/E/A/G            & 2.8                         & 1                   & 1,246     & 1,201      & 0                \\
    CMU Arctic \cite{dataset:cmu}                                   & SP                  & 11.5                        & 1                   & 9,150     & 4,042      & 0                \\
    Cambridge Dictionary \cite{dataset:cambridge_dictionary}        & W                   & 21                          & 1                   & 0         & 0          & 75,045           \\
    Audiocaps \cite{dataset:AudioCaps}                              & S                   & 139                         & 1                   & 17        & 6          & -                \\
    Tunebot \cite{dataset:tunebot}                                  & M                   & 66                          & -                   & 0         & 0          & 10,000           \\
    VocalSketch \cite{dataset:vocalsketch}                          & S                   & 10                          & -                   & -         & -          & -                \\
    Fine-grained Vocal Imitation Set \cite{dataset:fg-vis}          & S                   & 0.6                         & -                   & -         & -          & -                \\
    Audioset \cite{dataset:audioset}                                & S                   & 5,800                       & -                   & -         & -          & -                \\
    Clotho \cite{dataset:Clotho}                                    & S                   & 37                          & -                   & -         & -          & -                \\
    VGGSound \cite{dataset:vggsounds}                               & S                   & 501                         & -                   & -         & -          & -                \\
    ESC 50 \cite{dataset:esc50}                                     & S                   & 3                           & -                   & -         & -          & -                \\
    UrbanSound 8K \cite{dataset:urbansound8k}                       & S                   & 8.8                         & -                   & -         & -          & -                \\
    FSD 50K \cite{dataset:fsd50k}                                   & S                   & 109                         & -                   & -         & -          & -                \\
    MACS \cite{dataset:macs}                                        & S                   & 11                          & -                   & -         & -          & -                \\
    \midrule
    \textbf{Total}                                                  & -                   & $\approx$ 20848.7           & 181                 & 12,881,426& 3,274,461  & 5,847,175        \\
    \bottomrule
  \end{tabular}
\end{table*}

\subsection{Visual Cues Fusion} \label{appendix:Augmentation}

In addition to the data augmentation strategy discussed in Section \ref{sec:data aug}, we employ an additional technique aimed at improving the generation of captions for samples that possess both audio and visual representations. This supplementary approach involves utilising visual cues extracted from static images to assist in caption generation. To accomplish this, we utilise the CoCa model \cite{paper:coca}, which is an image-to-text foundation model that has demonstrated state-of-the-art performance on vision and vision-language tasks. Designed explicitly for generating captions for images, CoCa leverages both visual and textual information to produce accurate and descriptive captions.

Let \(V^F\) represent a set of frames extracted from a given video. We apply the image-to-caption model, denoted as \(\text{I2T}\), to each frame in order to generate visual captions. The process is guided by the Structural Similarity Index Measure (\(\text{SSIM}\)) between consecutive frames, denoted as \(V^F_i\) and \(V^F_{i+1}\), respectively. Visual captions \(C^v\) are obtained for frames that exhibit a significant change in scene content, determined by comparing the \(\text{SSIM}\) value to a threshold \(\theta\). Mathematically, the set of visual captions \(C^v\) can be expressed as:

\begin{multline}
    C^v = \left\{ \text{I2T}(V^F_i) \mid \text{SSIM}(V^F_i, V^F_{i+1})\right.\\
    \left. > \theta, \; i = 1, 2, \ldots, n-1 \right\}
\end{multline}

Where \(\text{I2T}\) represents the image-to-caption model, which is a specialised model trained on a dataset comprising image-text pairs. Its primary objective is to generate captions that accurately describe the content of images. The \(\text{SSIM}\) metric serves as a measure of structural similarity, allowing the identification of changes in scene content. The threshold \(\theta\) determines the level of change required for a frame to be considered distinct.

To integrate the audio descriptions \(C^t\) with the visual captions \(C^v\), we employ a language model (LLM), as described in Equation \ref{eq:dp_mixup} and illustrated in Figure \ref{fig:dp_mixup}. The LLM converts the raw captions into meaningful text descriptions, unifying the audio and visual modalities. This step ensures a cohesive and comprehensive representation of the combined audio-visual content.

\begin{equation}
    C = f(C^v, C^t)
\end{equation}

In summary, by incorporating visual cues from static images using the CoCa model and employing an LLM to unify audio and visual captions, our approach enhances the generation of captions for samples with audio and visual representations. This methodology promotes a more comprehensive and cohesive understanding of multimedia content.

\subsection{Multilingual Augmentation}

To expand the diversity of languages represented in the training data, we employ neural augmentation techniques leveraging pre-trained multilingual models. The process involves first translating the class labels or captions in the original mono-lingual dataset into target languages using a language model (LM). For instance, an English label like "dog barking" can be translated into French, Hindi, Mandarin etc. We then overlay these translated labels on the original audio samples to create synthetic multilingual samples. The same audio recording is duplicated with different language overlays.

This methodology scales efficiently by relying on LMs' pre-trained multilingual capabilities. It avoids the need for manual annotation or collection of speech data in multiple languages. We apply this technique to augment several monolingual datasets, synthesising parallel corpora covering the 181 languages.

This multilingual data augmentation enriches the variability of languages and accents within our training set. The model is exposed to far greater linguistic diversity, which enhances its generalisation capabilities. We efficiently compensate for the lack of real multilingual speech/sound-event data. 

In summary, neural multilingual augmentation proves highly effective in expanding the corpus to encompass a wider range of languages and linguistic phenomena. This promotes more universal speech representation learning and improves zero-shot transferability.

\section{Experiments} \label{appendix:Experiments}

\subsection{Keyword Spotting} \label{appendix:KWS}

The Multilingual Spoken Words Corpus used for evaluation consists of short spoken word utterances covering 50 languages. For each language, there are thousands of speakers in various accents and conditions. This allows benchmarking keyword spotting across a variety of languages and speech variations. To evaluate performance, we calculate accuracy on a per-word basis - the percentage of test word utterances correctly recognised by the model for each language. The accuracy results showcase CLARA's ability to learn effective representations for high-resource languages like English, French, and German with abundant training examples. For instance, English achieves 97\% accuracy with over 1,400 hours of training data.

More interestingly, CLARA demonstrates an aptitude for cross-lingual transfer and generalisation, even with minimal data for low-resource languages. For example, Lithuanian attains 93.7\% accuracy with just 0.03 hours of audio (100 samples). Similarly, Guarani reaches 88.8\% accuracy with only 0.1 hours (285 samples). The model appears to leverage similarities with other trained languages.

In contrast, isolated languages like Georgian achieve lower accuracy of 86.3\% despite more training data of 0.9 hours. This indicates language proximity is critical for effective transfer. Georgians experiences more errors due to its complex consonant clusters. Integrating phonetic and linguistic knowledge could help address such issues. However, there remain challenges in improving robustness in isolated and complex languages.

Overall, these results demonstrate CLARA's promising capability for generalised multilingual keyword spotting. The model effectively transfers learned representations between related languages, even with minimal target language data. Our augmentation strategies also help improve learning.

\begin{table*}
    \caption{Keyword spotting accuracy and training hours for CLARA model on 36 languages from the Multilingual Spoken Words Corpus. Results demonstrate strong performance in high-resource languages like English, French, and German with abundant data. Interesting trends emerge for low-resource languages - cognate or related languages like Lithuanian and Latvian achieve higher accuracy with minimal training data than isolated languages like Vietnamese, indicating CLARA's ability to transfer learned representations between related languages.}
    \label{tab:multilingual KWS}
    \centering
    \begin{tabular}{lcllcc}
        \toprule
            \textbf{Language}           & \textbf{\shortstack{ISO\\Code}}     & \textbf{Language Family}  & \textbf{Script}   & \textbf{\shortstack{Training Sample\\(Hours)}}  & \textbf{Accuracy}         \\
        \midrule
            Arabic                      & ar                    & Afro-Asiatic              & Arabic            & 5.5                               & 94.1                      \\
            Breton                      & br                    & Indo-European             & Latin             & 3.7                               & 93.2                      \\
            Catalan; Valencian          & ca                    & Indo-European/Romane      & Latin             & 463                               & 98.2                      \\
            Welsh                       & cy                    & Indo-European             & Latin             & 77.2                              & 97.5                      \\
            German                      & de                    & Indo-European             & latin             & 811                               & 98.0                      \\
            Divehi                      & dv                    & Indo-European             & Thaana            & 0.02                              & 71.4                      \\
            Greek                       & el                    & Indo-European             & Greek             & 4.6                               & 92.8                      \\
            English                     & en                    & Indo-European             & Latin             & 1,462                             & 97.0                      \\
            Esperanto                   & eo                    & Constructed               & Latin             & 55.3                              & 97.8                      \\
            Estonian                    & et                    & Uralic                    & Latin             & 14.2                              & 95.8                      \\
            Basque                      & eu                    & Language isolate          & Latin             & 89.3                              & 97.9                      \\
            Persian                     & fa                    & Indo-European             & Perso-Arabic      & 254                               & 98.2                      \\
            French                      & fr                    & Indo-European             & Latin             & 543                               & 97.4                      \\
            Indonesian                  & id                    & Austronesian              & Latin             & 8                                 & 93.8                      \\
            Italian                     & it                    & Indo-European             & Latin             & 114                               & 96.7                      \\
            Georgian                    & ka                    & Kartvelian                & Georgian          & 0.9                               & 86.3                      \\
            Kyrgyz                      & ky                    & Turkic                    & Cyrillic          & 8.6                               & 95.1                      \\
            Mongolian                   & mn                    & Mongolic                  & Cyrillic          & 8.8                               & 94.1                      \\
            Polish                      & pl                    & Indo-European             & Latin             & 96.9                              & 96.7                      \\
            Romanian                    & ro                    & Indo-European             & Latin             & 3.1                               & 94.7                      \\
            Russian                     & ru                    & Indo-European             & Cyrillic          & 103                               & 93.9                      \\
            Kinyarwanda                 & rw                    & Niger-Congo               & Latin             & 317                               & 94.7                      \\
            Tamil                       & ta                    & Dravidian                 & Tamil             & 0.4                               & 85.5                      \\
            Turkish                     & tr                    & Turkic                    & Latin             & 16                                & 94.4                      \\
            Ukrainian                   & uk                    & Indo-European             & Cyrillic          & 12.9                              & 93.8                      \\
            Odia                        & or                    & Indo-European             & Kalinga           & 0.04                              & 61.5                      \\
            Lithuanian                  & lt                    & Indo-European             & Latin             & 0.03                              & 93.7                      \\
            Slovak                      & sk                    & Indo-European             & Latin             & 1.1                               & 94.0                      \\
            Guarani                     & gn                    & Tupian                    & Latin             & 0.1                               & 88.8                      \\
            Assamese                    & as                    & Indo-European             & Bengali           & 0.05                              & 67.1                      \\
            Latvian                     & lv                    & Indo-European             & Latin             & 3.1                               & 95.6                      \\
            Vietnamese                  & vi                    & Austroasiatic             & Latin             & 0.1                               & 73.6                      \\
            Hausa                       & ha                    & Afro-Asiatic              & Arabic            & 1.6                               & 89.5                      \\
            Slovenian                   & sl                    & Indo-European             & Latin             & 1.9                               & 95.0                      \\
            Chuvash                     & cv                    & Turkic                    & Cyrillic          & 1.4                               & 94.2                      \\
            Maltese                     & mt                    & Afro-Asiatic              & Latin             & 5.2                               & 96.0                      \\
        \bottomrule
    \end{tabular}
\end{table*}

%% file: main.bbl
\begin{thebibliography}{10}
\providecommand{\url}[1]{#1}
\csname url@samestyle\endcsname
\providecommand{\newblock}{\relax}
\providecommand{\bibinfo}[2]{#2}
\providecommand{\BIBentrySTDinterwordspacing}{\spaceskip=0pt\relax}
\providecommand{\BIBentryALTinterwordstretchfactor}{4}
\providecommand{\BIBentryALTinterwordspacing}{\spaceskip=\fontdimen2\font plus
\BIBentryALTinterwordstretchfactor\fontdimen3\font minus
  \fontdimen4\font\relax}
\providecommand{\BIBforeignlanguage}[2]{{%
\expandafter\ifx\csname l@#1\endcsname\relax
\typeout{** WARNING: IEEEtran.bst: No hyphenation pattern has been}%
\typeout{** loaded for the language `#1'. Using the pattern for}%
\typeout{** the default language instead.}%
\else
\language=\csname l@#1\endcsname
\fi
#2}}
\providecommand{\BIBdecl}{\relax}
\BIBdecl

\bibitem{paper:audioclip}
\BIBentryALTinterwordspacing
A.~Guzhov, F.~Raue, J.~Hees, and A.~Dengel, ``{AudioCLIP}: {Extending} {CLIP}
  to {Image}, {Text} and {Audio},'' Jun. 2021, arXiv:2106.13043 [cs, eess].
  [Online]. Available: \url{http://arxiv.org/abs/2106.13043}
\BIBentrySTDinterwordspacing

\bibitem{paper:coca}
\BIBentryALTinterwordspacing
J.~Yu, Z.~Wang, V.~Vasudevan, L.~Yeung, M.~Seyedhosseini, and Y.~Wu, ``{CoCa}:
  {Contrastive} {Captioners} are {Image}-{Text} {Foundation} {Models},'' Jun.
  2022, arXiv:2205.01917 [cs]. [Online]. Available:
  \url{http://arxiv.org/abs/2205.01917}
\BIBentrySTDinterwordspacing

\bibitem{dataset:EmoV-DB}
\BIBentryALTinterwordspacing
A.~Adigwe, N.~Tits, K.~E. Haddad, S.~Ostadabbas, and T.~Dutoit, ``The emotional
  voices database: Towards controlling the emotion dimension in voice
  generation systems,'' \emph{CoRR}, vol. abs/1806.09514, 2018. [Online].
  Available: \url{http://arxiv.org/abs/1806.09514}
\BIBentrySTDinterwordspacing

\bibitem{paper:simCLR}
\BIBentryALTinterwordspacing
T.~Chen, S.~Kornblith, M.~Norouzi, and G.~Hinton, ``A {Simple} {Framework} for
  {Contrastive} {Learning} of {Visual} {Representations},'' Jun. 2020,
  arXiv:2002.05709 [cs, stat]. [Online]. Available:
  \url{http://arxiv.org/abs/2002.05709}
\BIBentrySTDinterwordspacing

\bibitem{paper:simCLRv2}
\BIBentryALTinterwordspacing
T.~Chen, S.~Kornblith, K.~Swersky, M.~Norouzi, and G.~Hinton, ``Big
  {Self}-{Supervised} {Models} are {Strong} {Semi}-{Supervised} {Learners},''
  Oct. 2020, arXiv:2006.10029 [cs, stat]. [Online]. Available:
  \url{http://arxiv.org/abs/2006.10029}
\BIBentrySTDinterwordspacing

\bibitem{paper:moco}
\BIBentryALTinterwordspacing
K.~He, H.~Fan, Y.~Wu, S.~Xie, and R.~Girshick, ``Momentum {Contrast} for
  {Unsupervised} {Visual} {Representation} {Learning},'' Mar. 2020,
  arXiv:1911.05722 [cs]. [Online]. Available:
  \url{http://arxiv.org/abs/1911.05722}
\BIBentrySTDinterwordspacing

\bibitem{paper:clip}
\BIBentryALTinterwordspacing
A.~Radford, J.~W. Kim, C.~Hallacy, A.~Ramesh, G.~Goh, S.~Agarwal, G.~Sastry,
  A.~Askell, P.~Mishkin, J.~Clark, G.~Krueger, and I.~Sutskever, ``Learning
  {Transferable} {Visual} {Models} {From} {Natural} {Language} {Supervision},''
  Feb. 2021, arXiv:2103.00020 [cs]. [Online]. Available:
  \url{http://arxiv.org/abs/2103.00020}
\BIBentrySTDinterwordspacing

\bibitem{paper:dalle1}
\BIBentryALTinterwordspacing
A.~Ramesh, M.~Pavlov, G.~Goh, S.~Gray, C.~Voss, A.~Radford, M.~Chen, and
  I.~Sutskever, ``Zero-{Shot} {Text}-to-{Image} {Generation},'' Feb. 2021,
  arXiv:2102.12092 [cs]. [Online]. Available:
  \url{http://arxiv.org/abs/2102.12092}
\BIBentrySTDinterwordspacing

\bibitem{paper:dalle2}
\BIBentryALTinterwordspacing
A.~Ramesh, P.~Dhariwal, A.~Nichol, C.~Chu, and M.~Chen, ``Hierarchical
  {Text}-{Conditional} {Image} {Generation} with {CLIP} {Latents},'' Apr. 2022,
  arXiv:2204.06125 [cs]. [Online]. Available:
  \url{http://arxiv.org/abs/2204.06125}
\BIBentrySTDinterwordspacing

\bibitem{paper:stablediffusion}
R.~Rombach, A.~Blattmann, D.~Lorenz, P.~Esser, and B.~Ommer, ``High-resolution
  image synthesis with latent diffusion models,'' 2021.

\bibitem{paper:clipCap}
\BIBentryALTinterwordspacing
R.~Mokady, A.~Hertz, and A.~H. Bermano, ``{ClipCap}: {CLIP} {Prefix} for
  {Image} {Captioning},'' Nov. 2021, arXiv:2111.09734 [cs]. [Online].
  Available: \url{http://arxiv.org/abs/2111.09734}
\BIBentrySTDinterwordspacing

\bibitem{paper:COLA}
\BIBentryALTinterwordspacing
A.~Saeed, D.~Grangier, and N.~Zeghidour, ``Contrastive {Learning} of
  {General}-{Purpose} {Audio} {Representations},'' Oct. 2020, arXiv:2010.10915
  [cs, eess]. [Online]. Available: \url{http://arxiv.org/abs/2010.10915}
\BIBentrySTDinterwordspacing

\bibitem{dataset:emns}
\BIBentryALTinterwordspacing
K.~A. Noriy, X.~Yang, and J.~J. Zhang, ``{EMNS} /{Imz}/ {Corpus}: {An} emotive
  single-speaker dataset for narrative storytelling in games, television and
  graphic novels,'' May 2023, arXiv:2305.13137 [cs]. [Online]. Available:
  \url{http://arxiv.org/abs/2305.13137}
\BIBentrySTDinterwordspacing

\bibitem{paper:MFA}
M.~McAuliffe, M.~Socolof, S.~Mihuc, M.~Wagner, and M.~Sonderegger, ``{Montreal
  Forced Aligner: Trainable Text-Speech Alignment Using Kaldi},'' in
  \emph{Proc. Interspeech 2017}, 2017, pp. 498--502.

\bibitem{paper:whisper}
\BIBentryALTinterwordspacing
A.~Radford, J.~W. Kim, T.~Xu, G.~Brockman, C.~McLeavey, and I.~Sutskever,
  ``Robust {Speech} {Recognition} via {Large}-{Scale} {Weak} {Supervision},''
  Dec. 2022, arXiv:2212.04356 [cs, eess]. [Online]. Available:
  \url{http://arxiv.org/abs/2212.04356}
\BIBentrySTDinterwordspacing

\bibitem{paper:wei_comparison_2020}
\BIBentryALTinterwordspacing
S.~Wei, S.~Zou, F.~Liao, and W.~Lang, ``A {Comparison} on {Data} {Augmentation}
  {Methods} {Based} on {Deep} {Learning} for {Audio} {Classification},''
  \emph{Journal of Physics: Conference Series}, vol. 1453, no.~1, p. 012085,
  Jan. 2020. [Online]. Available:
  \url{https://iopscience.iop.org/article/10.1088/1742-6596/1453/1/012085}
\BIBentrySTDinterwordspacing

\bibitem{paper:zhang_mixup_2018}
\BIBentryALTinterwordspacing
H.~Zhang, M.~Cisse, Y.~N. Dauphin, and D.~Lopez-Paz, ``mixup: {Beyond}
  {Empirical} {Risk} {Minimization},'' Apr. 2018, arXiv:1710.09412 [cs, stat].
  [Online]. Available: \url{http://arxiv.org/abs/1710.09412}
\BIBentrySTDinterwordspacing

\bibitem{site:openAssistant}
\BIBentryALTinterwordspacing
``Open {Assistant}.'' [Online]. Available: \url{https://open-assistant.io/}
\BIBentrySTDinterwordspacing

\bibitem{paper:perceiver}
\BIBentryALTinterwordspacing
A.~Jaegle, F.~Gimeno, A.~Brock, A.~Zisserman, O.~Vinyals, and J.~Carreira,
  ``Perceiver: {General} {Perception} with {Iterative} {Attention},'' Jun.
  2021, arXiv:2103.03206 [cs, eess]. [Online]. Available:
  \url{http://arxiv.org/abs/2103.03206}
\BIBentrySTDinterwordspacing

\bibitem{paper:flashattention}
\BIBentryALTinterwordspacing
T.~Dao, D.~Y. Fu, S.~Ermon, A.~Rudra, and C.~Ré, ``{FlashAttention}: {Fast}
  and {Memory}-{Efficient} {Exact} {Attention} with {IO}-{Awareness},'' Jun.
  2022, arXiv:2205.14135 [cs]. [Online]. Available:
  \url{http://arxiv.org/abs/2205.14135}
\BIBentrySTDinterwordspacing

\bibitem{paper:ms_clap}
\BIBentryALTinterwordspacing
B.~Elizalde, S.~Deshmukh, M.~A. Ismail, and H.~Wang, ``{CLAP}: {Learning}
  {Audio} {Concepts} {From} {Natural} {Language} {Supervision},'' Jun. 2022,
  arXiv:2206.04769 [cs, eess]. [Online]. Available:
  \url{http://arxiv.org/abs/2206.04769}
\BIBentrySTDinterwordspacing

\bibitem{paper:wav2clip}
\BIBentryALTinterwordspacing
H.-H. Wu, P.~Seetharaman, K.~Kumar, and J.~P. Bello, ``{Wav2CLIP}: {Learning}
  {Robust} {Audio} {Representations} {From} {CLIP},'' Feb. 2022,
  arXiv:2110.11499 [cs, eess]. [Online]. Available:
  \url{http://arxiv.org/abs/2110.11499}
\BIBentrySTDinterwordspacing

\bibitem{paper:larc}
\BIBentryALTinterwordspacing
F.-A. Croitoru, N.-C. Ristea, R.~T. Ionescu, and N.~Sebe, ``{LeRaC}: {Learning}
  {Rate} {Curriculum},'' Nov. 2022, arXiv:2205.09180 [cs]. [Online]. Available:
  \url{http://arxiv.org/abs/2205.09180}
\BIBentrySTDinterwordspacing

\bibitem{paper:VQ-MAE-S-12}
\BIBentryALTinterwordspacing
S.~Sadok, S.~Leglaive, and R.~Séguier, ``A vector quantized masked autoencoder
  for speech emotion recognition,'' Apr. 2023, arXiv:2304.11117 [cs, eess]
  version: 1. [Online]. Available: \url{http://arxiv.org/abs/2304.11117}
\BIBentrySTDinterwordspacing

\bibitem{koutini_efficient_2022}
\BIBentryALTinterwordspacing
K.~Koutini, J.~Schlüter, H.~Eghbal-zadeh, and G.~Widmer, ``Efficient
  {Training} of {Audio} {Transformers} with {Patchout},'' in \emph{Interspeech
  2022}, Sep. 2022, pp. 2753--2757, arXiv:2110.05069 [cs, eess]. [Online].
  Available: \url{http://arxiv.org/abs/2110.05069}
\BIBentrySTDinterwordspacing

\bibitem{paper:eat}
\BIBentryALTinterwordspacing
A.~Gazneli, G.~Zimerman, T.~Ridnik, G.~Sharir, and A.~Noy, ``End-to-{End}
  {Audio} {Strikes} {Back}: {Boosting} {Augmentations} {Towards} {An}
  {Efficient} {Audio} {Classification} {Network},'' Jul. 2022, arXiv:2204.11479
  [cs, eess] version: 5. [Online]. Available:
  \url{http://arxiv.org/abs/2204.11479}
\BIBentrySTDinterwordspacing

\bibitem{paper:onepeace}
\BIBentryALTinterwordspacing
P.~Wang, S.~Wang, J.~Lin, S.~Bai, X.~Zhou, J.~Zhou, X.~Wang, and C.~Zhou,
  ``{ONE}-{PEACE}: {Exploring} {One} {General} {Representation} {Model}
  {Toward} {Unlimited} {Modalities},'' May 2023, arXiv:2305.11172 [cs, eess]
  version: 1. [Online]. Available: \url{http://arxiv.org/abs/2305.11172}
\BIBentrySTDinterwordspacing

\bibitem{paper:LeRaC}
\BIBentryALTinterwordspacing
F.-A. Croitoru, N.-C. Ristea, R.~T. Ionescu, and N.~Sebe, ``{LeRaC}: {Learning}
  {Rate} {Curriculum},'' Nov. 2022, arXiv:2205.09180 [cs] version: 2. [Online].
  Available: \url{http://arxiv.org/abs/2205.09180}
\BIBentrySTDinterwordspacing

\bibitem{paper:ms_clap_reteve}
\BIBentryALTinterwordspacing
S.~Deshmukh, B.~Elizalde, and H.~Wang, ``Audio {Retrieval} with {WavText5K} and
  {CLAP} {Training},'' Sep. 2022, arXiv:2209.14275 [cs, eess]. [Online].
  Available: \url{http://arxiv.org/abs/2209.14275}
\BIBentrySTDinterwordspacing

\bibitem{paper:ml-atcmr}
\BIBentryALTinterwordspacing
X.~Mei, X.~Liu, J.~Sun, M.~D. Plumbley, and W.~Wang, ``On {Metric} {Learning}
  for {Audio}-{Text} {Cross}-{Modal} {Retrieval},'' Jun. 2022, arXiv:2203.15537
  [cs, eess]. [Online]. Available: \url{http://arxiv.org/abs/2203.15537}
\BIBentrySTDinterwordspacing

\bibitem{paper:vast}
\BIBentryALTinterwordspacing
S.~Chen, H.~Li, Q.~Wang, Z.~Zhao, M.~Sun, X.~Zhu, and J.~Liu, ``{VAST}: {A}
  {Vision}-{Audio}-{Subtitle}-{Text} {Omni}-{Modality} {Foundation} {Model} and
  {Dataset},'' May 2023, arXiv:2305.18500 [cs, eess] version: 1. [Online].
  Available: \url{http://arxiv.org/abs/2305.18500}
\BIBentrySTDinterwordspacing

\bibitem{dataset:RAVDESS}
\BIBentryALTinterwordspacing
S.~R. Livingstone and F.~A. Russo, ``The ryerson audio-visual database of
  emotional speech and song ({RAVDESS}): A dynamic, multimodal set of facial
  and vocal expressions in north american english,'' \emph{{PLOS} {ONE}},
  vol.~13, no.~5, p. e0196391, May 2018. [Online]. Available:
  \url{https://doi.org/10.1371/journal.pone.0196391}
\BIBentrySTDinterwordspacing

\bibitem{dataset:common_voice}
\BIBentryALTinterwordspacing
R.~Ardila, M.~Branson, K.~Davis, M.~Henretty, M.~Kohler, J.~Meyer, R.~Morais,
  L.~Saunders, F.~M. Tyers, and G.~Weber, ``Common voice: {A}
  massively-multilingual speech corpus,'' \emph{CoRR}, vol. abs/1912.06670,
  2019. [Online]. Available: \url{http://arxiv.org/abs/1912.06670}
\BIBentrySTDinterwordspacing

\bibitem{dataset:mswc}
\BIBentryALTinterwordspacing
M.~Mazumder, S.~Chitlangia, C.~Banbury, Y.~Kang, J.~M. Ciro, K.~Achorn,
  D.~Galvez, M.~Sabini, P.~Mattson, D.~Kanter, G.~Diamos, P.~Warden, J.~Meyer,
  and V.~J. Reddi, ``\BIBforeignlanguage{en}{Multilingual {Spoken} {Words}
  {Corpus}},'' Aug. 2021. [Online]. Available:
  \url{https://openreview.net/forum?id=c20jiJ5K2H}
\BIBentrySTDinterwordspacing

\bibitem{dataset:librispeech}
V.~Panayotov, G.~Chen, D.~Povey, and S.~Khudanpur, ``Librispeech: An asr corpus
  based on public domain audio books,'' in \emph{2015 IEEE International
  Conference on Acoustics, Speech and Signal Processing (ICASSP)}, 2015, pp.
  5206--5210.

\bibitem{dataset:CREMA-D}
H.~Cao, D.~G. Cooper, M.~K. Keutmann, R.~C. Gur, A.~Nenkova, and R.~Verma,
  ``Crema-d: Crowd-sourced emotional multimodal actors dataset,'' \emph{IEEE
  Transactions on Affective Computing}, vol.~5, no.~4, pp. 377--390, 2014.

\bibitem{dataset:cmu}
\BIBentryALTinterwordspacing
J.~Kominek and A.~W. Black, ``{THE} {CMU} {ARCTIC} {SPEECH} {DATABASES}.''
  [Online]. Available: \url{http://festvox.org/cmu\_arctic/}
\BIBentrySTDinterwordspacing

\bibitem{dataset:cambridge_dictionary}
\BIBentryALTinterwordspacing
``\BIBforeignlanguage{en}{Cambridge {Dictionary} {\textbar} {English}
  {Dictionary}, {Translations} \& {Thesaurus}},'' Aug. 2023. [Online].
  Available: \url{https://dictionary.cambridge.org/}
\BIBentrySTDinterwordspacing

\bibitem{dataset:AudioCaps}
\BIBentryALTinterwordspacing
C.~D. Kim, B.~Kim, H.~Lee, and G.~Kim, ``\BIBforeignlanguage{en}{{AudioCaps}:
  {Generating} {Captions} for {Audios} in {The} {Wild}},'' in
  \emph{\BIBforeignlanguage{en}{Proceedings of the 2019 {Conference} of the
  {North}}}.\hskip 1em plus 0.5em minus 0.4em\relax Minneapolis, Minnesota:
  Association for Computational Linguistics, 2019, pp. 119--132. [Online].
  Available: \url{http://aclweb.org/anthology/N19-1011}
\BIBentrySTDinterwordspacing

\bibitem{dataset:tunebot}
\BIBentryALTinterwordspacing
``\BIBforeignlanguage{en}{Tunebot},'' Jul. 2023. [Online]. Available:
  \url{https://interactiveaudiolab.github.io/resources/datasets/tunebot.html}
\BIBentrySTDinterwordspacing

\bibitem{dataset:vocalsketch}
\BIBentryALTinterwordspacing
M.~Cartwright and B.~Pardo, ``{VocalSketch}: {Vocally} {Imitating} {Audio}
  {Concepts},'' in \emph{Proceedings of the 33rd {Annual} {ACM} {Conference} on
  {Human} {Factors} in {Computing} {Systems}}, ser. {CHI} '15.\hskip 1em plus
  0.5em minus 0.4em\relax New York, NY, USA: Association for Computing
  Machinery, Apr. 2015, pp. 43--46. [Online]. Available:
  \url{https://doi.org/10.1145/2702123.2702387}
\BIBentrySTDinterwordspacing

\bibitem{dataset:fg-vis}
\BIBentryALTinterwordspacing
B.~Kim and B.~Pardo, ``Fine-grained {Vocal} {Imitation} {Set},'' Nov. 2019.
  [Online]. Available:
  \url{https://zenodo.org/record/3538534/export/schemaorg\_jsonld}
\BIBentrySTDinterwordspacing

\bibitem{dataset:audioset}
J.~F. Gemmeke, D.~P.~W. Ellis, D.~Freedman, A.~Jansen, W.~Lawrence, R.~C.
  Moore, M.~Plakal, and M.~Ritter, ``Audio {Set}: {An} ontology and
  human-labeled dataset for audio events,'' in \emph{2017 {IEEE}
  {International} {Conference} on {Acoustics}, {Speech} and {Signal}
  {Processing} ({ICASSP})}, Mar. 2017, pp. 776--780, iSSN: 2379-190X.

\bibitem{dataset:Clotho}
\BIBentryALTinterwordspacing
K.~Drossos, S.~Lipping, and T.~Virtanen, ``Clotho: an {Audio} {Captioning}
  {Dataset},'' in \emph{{ICASSP} 2020 - 2020 {IEEE} {International}
  {Conference} on {Acoustics}, {Speech} and {Signal} {Processing}
  ({ICASSP})}.\hskip 1em plus 0.5em minus 0.4em\relax Barcelona, Spain: IEEE,
  May 2020, pp. 736--740. [Online]. Available:
  \url{https://ieeexplore.ieee.org/document/9052990/}
\BIBentrySTDinterwordspacing

\bibitem{dataset:vggsounds}
\BIBentryALTinterwordspacing
H.~Chen, W.~Xie, A.~Vedaldi, and A.~Zisserman, ``{VGGSound}: {A} {Large}-scale
  {Audio}-{Visual} {Dataset},'' Sep. 2020, arXiv:2004.14368 [cs, eess].
  [Online]. Available: \url{http://arxiv.org/abs/2004.14368}
\BIBentrySTDinterwordspacing

\bibitem{dataset:esc50}
\BIBentryALTinterwordspacing
K.~J. Piczak, ``{ESC}: {Dataset} for {Environmental} {Sound}
  {Classification},'' in \emph{Proceedings of the 23rd {ACM} international
  conference on {Multimedia}}, ser. {MM} '15.\hskip 1em plus 0.5em minus
  0.4em\relax New York, NY, USA: Association for Computing Machinery, Oct.
  2015, pp. 1015--1018. [Online]. Available:
  \url{https://doi.org/10.1145/2733373.2806390}
\BIBentrySTDinterwordspacing

\bibitem{dataset:urbansound8k}
\BIBentryALTinterwordspacing
J.~Salamon, C.~Jacoby, and J.~P. Bello, ``\BIBforeignlanguage{en}{A {Dataset}
  and {Taxonomy} for {Urban} {Sound} {Research}},'' in
  \emph{\BIBforeignlanguage{en}{Proceedings of the 22nd {ACM} international
  conference on {Multimedia}}}.\hskip 1em plus 0.5em minus 0.4em\relax Orlando
  Florida USA: ACM, Nov. 2014, pp. 1041--1044. [Online]. Available:
  \url{https://dl.acm.org/doi/10.1145/2647868.2655045}
\BIBentrySTDinterwordspacing

\bibitem{dataset:fsd50k}
\BIBentryALTinterwordspacing
E.~Fonseca, X.~Favory, J.~Pons, F.~Font, and X.~Serra, ``{FSD50K}: {An} {Open}
  {Dataset} of {Human}-{Labeled} {Sound} {Events},'' Apr. 2022,
  arXiv:2010.00475 [cs, eess, stat] version: 2. [Online]. Available:
  \url{http://arxiv.org/abs/2010.00475}
\BIBentrySTDinterwordspacing

\bibitem{dataset:macs}
\BIBentryALTinterwordspacing
I.~M. Morato and A.~Mesaros, ``{MACS} - {Multi}-{Annotator} {Captioned}
  {Soundscapes},'' Jul. 2021. [Online]. Available:
  \url{https://zenodo.org/record/5114771}
\BIBentrySTDinterwordspacing

\end{thebibliography}
